\documentclass[review]{iopjournal}

\hypersetup{colorlinks=true,linkcolor=blue,citecolor=blue,urlcolor=blue}

\usepackage{graphicx}% Include figure files
\usepackage{dcolumn}% Align table columns on decimal point
\usepackage{bm}% bold math
\usepackage{amsmath}
\usepackage{xcolor}
\usepackage{float}
\usepackage{subcaption}
\usepackage[normalem]{ulem}

\usepackage[mathlines]{lineno}% Enable numbering of text and display math
\usepackage[numbers,sort&compress]{natbib}

\usepackage{setspace}

\begin{document}
%\doublespacing

\articletype{Paper} %     e.g. Paper, Letter, Topical Review...

\title{Non-Ideal Relativistic Plasma Mirrors \footnote{This paper is dedicated to Prof. Tito Mendonça on his 80th birthday for his continuous inspiration and source of novel ideas}}

\author{Zahra M. Chamazkoti$^{*}$\orcid{0000-0002-7157-615X} and Luís O. Silva\orcid{0000-0003-2906-924X}}

\affil{GoLP/Instituto de Plasmas e Fusão Nuclear, Instituto Superior Técnico,
Universidade de Lisboa,  1049-001 Lisbon, Portugal}

\affil{$^*$Author to whom any correspondence should be addressed.}

\email{zahra.chamazkoti@tecnico.ulisboa.pt}


\begin{abstract}
Relativistic plasma mirrors formed in the wake of intense laser pulses propagating in underdense plasmas can provide an efficient mechanism for generating attosecond pulses, and for enhanced focusing towards ultra-high intensity fields. We derive the reflection coefficient of a laser pulse interacting with relativistic mirrors, assuming different density profiles for the mirror. Analytical expressions are obtained for flat-top and parabolic density mirrors, and in the Dirac-delta limit, demonstrating the strong dependence of the reflectivity on the plasma mirror density profile. Fully kinetic multi-dimensional particle-in-cell simulations and our analytical model confirm that the density spike at the rear of the plasma bubble exhibits an approximately parabolic profile. Our results show that a laser wakefield setup can be optimized to maximize the reflection coefficient under specific laser driver conditions, determined by the balance between the moving plasma mirror proper density, thickness, and relativistic momentum carried by the mirror.

\end{abstract}

\section{Introduction}

The generation of attosecond pulses~\cite{corkum2000attosecond, krausz2009attosecond} has opened new regimes in the interaction of intense radiation with matter. These pulses are typically produced through the coherent buildup of high harmonics during nonlinear laser propagation in gases~\cite{l1993high, paul2001observation} or at relativistically oscillating plasma surfaces formed on overdense solid targets~\cite{thaury2010high,d2006route,jahn2019towards}. 
Proposed plasma-based compression mechanisms for laser pulses include the self-modulation of ultra-intense laser pulses~\cite{tsung2002generation}, co-propagation with relativistic plasma waves~\cite{wilks1988frequency, sheng2000generation}, and interaction with ionization fronts~\cite{kapteyn1991relativistic}. Ionization fronts can act as relativistic mirrors~\cite{LAMPE.1978}, but reflection is efficient only near the Doppler-shifted cut-off and is generally more effective in transmission~\cite{mori1991generation, mendoncca1994regular}. Relativistic mirrors based on compressed electron beams~\cite{landecker1952possibility, granatstein1976realization} are limited to microwave frequencies, where reflection is efficient, as current beams do not reach the densities required for significant reflection at optical and near-infrared frequencies~\cite{mendoncca2000theory}. 

In this work, we investigate the reflection of a source laser pulse from the relativistic nonlinear wakefield driven by an ultra-intense laser in underdense plasma~\cite{wilks1988frequency, decker1996evolution}. This concept was first explored by Decker \textit{et al.}~\cite{decker1996evolution}, generalized by Bulanov \textit{et al.}~\cite{bulanov2003light} as a mechanism for approaching the Schwinger limit—the critical electric field scale for vacuum electron–positron pair production~\cite{schwinger1951gauge}—and later extended to realistic implementations of curved relativistic mirrors based on optically curved plasma mirrors~\cite{vincenti2019achieving,kato2026high}. Unlike relativistic ionization fronts, the density spike formed behind the driver carries finite momentum and can partially reflect a counter-propagating source pulse. The reflected pulse is frequency upshifted and temporally compressed by the relativistic double Doppler effect, since the number of oscillations is Lorentz invariant. The combination of high density and relativistic motion yields significantly higher reflectivity for ultrashort pulse generation than other proposed mechanisms~\cite{bulanov2003light, naumova2004relativistic, gordienko2004relativistic}. For a review of analytical methods for relativistic mirrors in laser plasmas, see~\cite{bulanov2016relativistic}. Relativistic plasma mirrors have also been proposed as laboratory platforms for analog black-hole physics~\cite{chen2017accelerating}.

Following J.~E.~Santos \textit{et al.}~\cite{santos2002attosecond,santos2003relativistic}, we consider different mirror density profiles to derive the reflectivity of the relativistic plasma mirror. 
Section~2 reviews the main properties of the relativistic density spike driven in an underdense plasma, within the quasi-static approximation, following~\cite{martins2004pulse}. In section~3, we derive the reflection and transmission coefficients of an idealized relativistic plasma mirror, and for different model density profiles, generalizing previous work~\cite{santos2002attosecond,santos2003relativistic}. In section~4, we extend the model to laser-driven mirrors and validate the theory with multi-dimensional particle-in-cell simulations using \textsc{Osiris}~\cite{fonseca2002lecture}. Finally, we state the conclusions.

\section{Laser-Driven Plasma Mirrors} 
In our configuration, an intense laser driver pulse propagates through an underdense plasma, exciting a strongly nonlinear wakefield. Martins \textit{et al.}~\cite{martins2004pulse} employed a 1D quasi-static model of nonlinear laser--plasma interaction~\cite{sprangle1990nonlinear} to describe the formation of a density spike at the rear of the wakefield. In this model, the laser driver is represented by a step-like profile
$a_d(\xi)=a_0[\Theta(\xi+L_d)-\Theta(\xi)]$,
with $\xi=k_{p0}(x-ct)$~\cite{fonseca2002lecture}, where $\Theta$ is the Heaviside function and $a_0= eE_0 / m_e \omega_d c$ is the dimensionless peak vector potential of the driver pulse. $\omega_d$ is the driver pulse frequency and $k_{p0}=\omega_{p0}/c$, where $\omega_{p0} = (4\pi e^2 n_{e0}/m_e)^{1/2}$ is the background plasma frequency for density $n_{e0}$.
The optimal pulse length that maximizes the longitudinal wakefield excitation, corresponding to the maximum energy transfer from the laser pulse to the plasma wave, is
$L_d = 2\sqrt{1+a_0^2}\; E\!\left(\sqrt{{a_0^2} / {1+a_0^2}}\right)$,
where $E(x)$ is the complete elliptic integral of the second kind~\cite{martins2004pulse,berezhiani1990relativistic,olver2010nist,byrd1971handbook}. Within this model, the electron density and Lorentz factor profiles are given by
\begin{equation}
n_e(\xi)
=
\frac{1}{2}
\left[
1+\frac{1+a^2(\xi)}{y^2(\xi)}
\right],
\label{eq:density_profile_y}
\end{equation}
and
\begin{equation}
\gamma_e(\xi)
=
\frac{1}{2}
\left[
y(\xi)+\frac{1+a^2(\xi)}{y(\xi)}
\right].
\label{eq:gamma_profile_y}
\end{equation}
Here, $y(\xi)=1+\Phi(\xi)$, $\Phi(\xi)=e\phi(\xi)/m_ec^2$, where $\phi$ is the electrostatic wake potential. Along the wake branch extending from the rear edge of the driver to the density maximum, $y(\xi)$ decreases from $\gamma_\perp^2(\xi)$ to $\gamma_\perp^{-2}(\xi)$, where $\gamma_\perp=(1+a^2(\xi))^{1/2}$. For the step-like driver considered here, $a(\xi)=a_0$ inside the pulse and $a(\xi)=0$ behind it. Using equations~\eqref{eq:density_profile_y} and~\eqref{eq:gamma_profile_y}, together with the quasi-static wake solution, the maximum density of the density spike $n_{\mathrm{e\max}}$, the maximum Lorentz factor of the spike $\gamma_{\max}$, and the spike thickness $l_{\mathrm{mirror}}$ are given by

\begin{equation}
n_{\mathrm{e\max}}=\tfrac{1}{2}\left(1+\gamma_\perp^4\right),
\label{nemax}
\end{equation}

\begin{equation}
\gamma_{\max}=\frac{n_{\mathrm{e\max}}}{\gamma_\perp^2},
\label{gammamax}
\end{equation}

\begin{equation}
l_{\text{mirror}} = 4\gamma_{\perp} \text{Im}\left[E\left(\arcsin\, \gamma_{\perp} \big| \tfrac{1}{\gamma_{\perp}^4}\right) - F\left(\arcsin \, \gamma_{\perp} \big| \tfrac{1}{\gamma_{\perp}^4}\right)\right],
\label{lmirror}
\end{equation}
where $F(\varphi|m)$ and $E(\varphi|m)$ denote the elliptic integrals of the first and second kind~\cite{abramowitz1965handbook,olver2010nist,byrd1971handbook}.
In this paper, and unless explicitly stated, lengths are normalized to $c/\omega_{p0}$, time to $\omega_{p0}^{-1}$, and mass and charge to $m_e$ and $e$, respectively. Equations~\eqref{nemax}--\eqref{lmirror} show that, in the quasi-static approximation, $n_{\mathrm{e\max}}$, $\gamma_{\max}$, and $l_{\mathrm{mirror}}$ depend only on the driver laser intensity.

We also note the dependence of the upshifted frequency on $a_0$ in this model. 
Using equations~\eqref{nemax} and~\eqref{gammamax}, together with the standard double-Doppler upshift for normal incidence on a relativistic mirror~\cite{einstein1905electrodynamics}, the maximum upshift for a relativistic plasma mirror becomes $\omega_r =
\left[(1+a_0^2) + {1} / {(1+a_0^2)}\right]^2 \omega_0$. 
This expression gives $\omega_r \approx 4\,\omega_0$ in the non-relativistic limit $(a_0 \ll 1)$, and 
$\omega_r \approx a_0^4\,\omega_0$ in the ultra-relativistic limit $(a_0 \gg 1)$.

\begin{figure}[H]
 \centering
\includegraphics[width=0.65\linewidth,height=12.450cm]{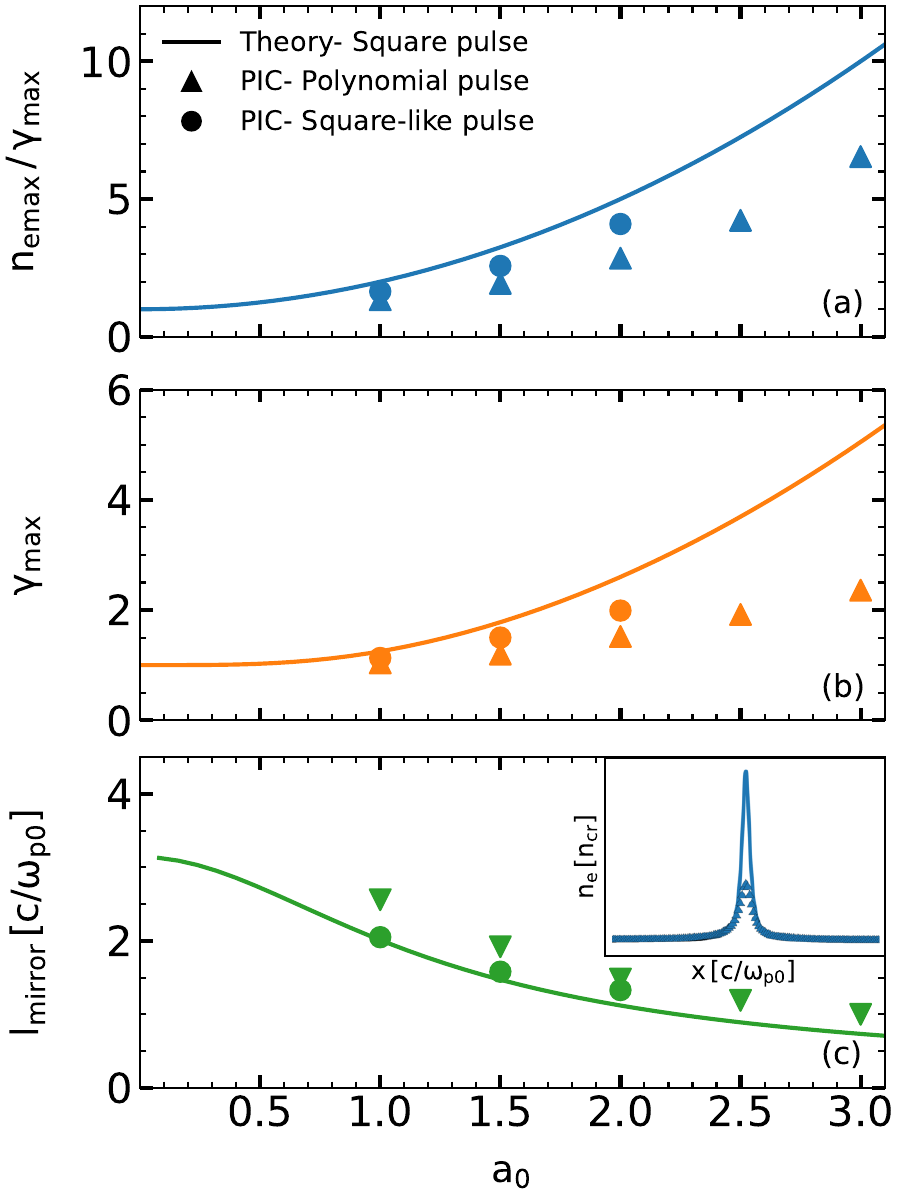}
\caption{Comparison of the plasma-mirror parameters (a) $n_{\mathrm{e\max}}/\gamma_{\max}$, (b) $\gamma_{\max}$, and (c) $l_{\rm mirror}$ as functions of $a_0$ between 1D quasi-static theoretical predictions using an ideal square driver (curves) and 2D \textsc{Osiris} simulations (symbols). Triangles correspond to simulations with a smooth polynomial (Gaussian-like) driver, while circles denote square-like (ramp–flat–ramp) pulses with polynomial ramps. The simulations follow the analytical trends and remain of the same order, with quantitative differences arising from the lower energy, $\int_{-\infty}^{+\infty} a^2(\xi)\,d\xi$, of the polynomial pulse relative to the ideal square driver, as well as from pulse shape and multidimensional effects. The inset in (c) compares a lineout of the electron density spike at the rear of the simulated plasma bubble with the analytical model given by equation~\eqref{eq:density_profile_y}.}

\label{ngl}
\end{figure}

In figure~\ref{ngl}, we compare the 1D quasi-static predictions (curves) with \textsc{Osiris} simulations (circles and triangles). The analytical model assumes an ideal step-like (square) driver, whereas the simulations employ square-like (ramp--flat--ramp) and Gaussian-like (polynomial) temporal envelopes~\cite{fonseca2002lecture} (see Supplementary data).
For the same peak amplitude, and taking $L_d$ and $2L_d$ as the effective durations of the square and polynomial pulses, respectively, the polynomial driver carries $\simeq78\%$ of the square pulse energy ($\int_{-\infty}^{+\infty} a^2(\xi)d\xi$). This partly explains the quantitative discrepancy between the curves and triangle points in figure~\ref{ngl}, while preserving the overall trends.

To isolate pulse shape effects beyond energy mismatch, we performed additional simulations using a square-like temporal profile with polynomial ramps and effective duration $L_d$, with energy matched to the polynomial pulse.
The resulting data (circles in figure~\ref{ngl}) move closer to the quasi-static predictions but do not fully coincide, indicating residual sensitivity to the driver temporal shape. Furthermore, for $a_0 \ge 2.5$ with the square-like driver pulse, we observe self-injection in our simulations, and the density spike deforms before a well-defined mirror forms, preventing an unambiguous extraction of $n_{\mathrm{e\max}}$, $\gamma_{\max}$, and $l_{\rm mirror}$. Finally, the analytical model is strictly one-dimensional (1D), whereas the simulations are multi-dimensional, which can also explain the discrepancy. All simulation parameters are provided in the Supplementary data.

Our simulations and the analytical model in equation~\eqref{eq:density_profile_y} also make evident the finite thickness of the mirror $l_{\mathrm{mirror}}$ and the non-ideal nature of the density spike $n_e$ (see inset in figure~\ref{ngl}). This motivates a more in-depth study of the electromagnetic properties of these structures, in terms of reflection and transmission.

\begin{figure}
 \centering
\includegraphics[width=0.85\linewidth,height=10.0cm]{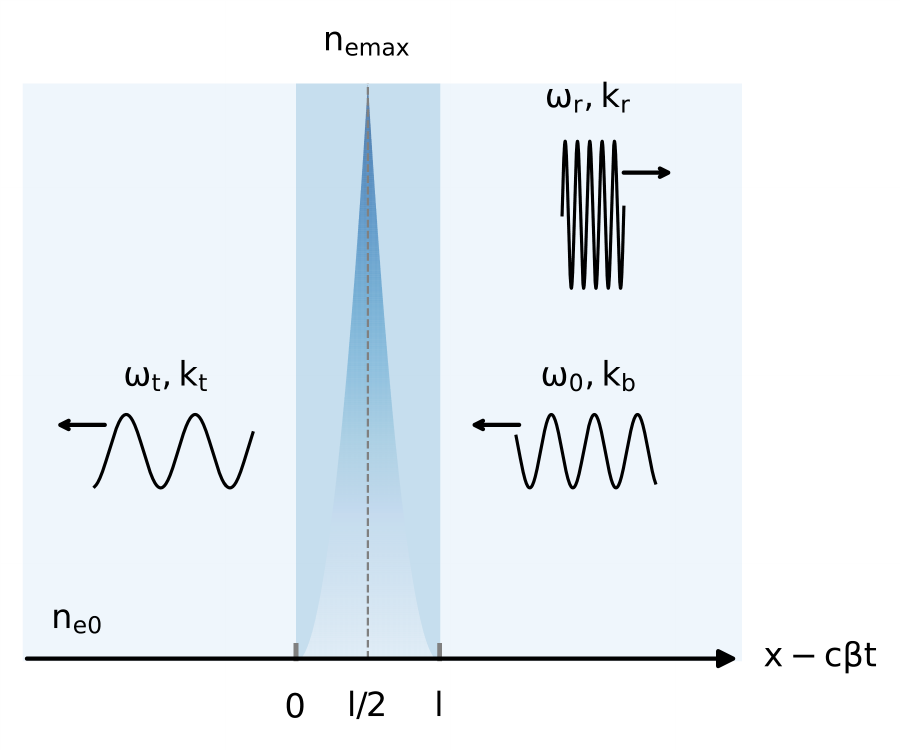}
\caption{Schematic of the 1D configuration used in the analytical model, shown in the laboratory frame. A linearly $z$-polarized source pulse counter-propagates along $x$ and interacts with a relativistic density structure embedded in a uniform background of density $n_{e0}$. The mirror is moving from left to right and occupies $0<x<l$ and is modeled either as a flat-top slab with $n_e (x)=n_{\mathrm{e\max}}$ and as a parabolic spike that rises from $n_{e0}$ at both edges of the mirror to $n_{\mathrm{e\max}}$ at $x=l/2$.}
\label{setup}
\end{figure}

\section{Reflection of electromagnetic pulses beyond idealized plasma mirrors}

We now derive the reflection and transmission coefficients of a moving mirror for three density profiles—flat-top, parabolic, and in the Dirac-$\delta$ limit.
A realistic relativistic plasma mirror is a non-ideal mirror with finite thickness and a specific density profile, as shown in the inset of figure~\ref{ngl}(c). The piecewise parabolic profile is chosen because it is analytically tractable. Our fits to the 2D \textsc{Osiris} data show that it provides a good approximation to the density spike at the rear of the plasma bubble, particularly near its maximum.
 
In the mirror rest frame, under these conditions, the reflectivity is modified and below unity, in contrast to the idealized infinitely thin, delta-like mirror~\cite{bulanov2003light}. The Dirac-$\delta$ limit provides an analytically tractable ultrathin-mirror model that removes thickness-induced interference effects and isolates the role of the mirror areal charge, providing a clear benchmark for comparison with realistic, narrow wakefield density spikes. As we will see, this idealized model significantly overestimates the reflectivity of the mirror.

As shown in figure~\ref{setup}, a linearly polarized source pulse with frequency $\omega_0$ counter-propagates along $x$ and collides with the moving mirror, leading to partial reflection from this dense relativistic electron layer, producing a blueshifted and temporally compressed pulse~\cite{LAMPE.1978}. In the proper frame, the mirror is stationary and occupies the region $0<x'<l'$ in the background plasma with density  $n'_e (x') =n'_{e0}$. We note that the transformation from the proper frame to the lab frame follows $l'=\gamma\,l$ and $n'_e=n_e/\gamma$, where $l =  l_{\mathrm{mirror}}$ is the lab-frame mirror thickness given by equation~\eqref{lmirror}, $\gamma=(1-\beta^2)^{-1/2}$, and $\beta$ is the normalized velocity of the mirror in the lab frame. For a flat-top profile, the mirror electron density in the proper frame is uniform, $n'_e(x')=n'_{\mathrm{e\max}}$, whereas for a parabolic profile, $n'_e(x')$ increases smoothly from $n'_{e0}$ at both edges of the mirror to $n'_{\mathrm{e\max}}$ at $x'=l'/2$, with $n'_e(x')=n'_{e0}+(n'_{\mathrm{e\max}}-n'_{e0})(2x'/l')^2$ for $0<x'<l'/2$, and $n'_e(x')=n'_{e0}+(n'_{\mathrm{e\max}}-n'_{e0})(2(x'-l')/l')^2$ for $l'/2<x'<l'$.
 All primed quantities refer to the proper frame, while unprimed quantities refer to the laboratory frame.

The reflection and transmission coefficients of moving mirrors are derived following a standard procedure in the proper frame~\cite{LAMPE.1978} and then transformed to the laboratory frame. We follow closely the techniques described in Born and Wolf~\cite{born2013principles} for stratified media. 
Assuming time-harmonic field $E'(x',t')=e^{-i\omega_0' t'}E'(x')$, the Helmholtz equation for electromagnetic wave propagation reads

\begin{equation}
\frac{d^2E'}{dx'^2}+\frac{\big[\omega_0'^2-\omega_p^2(x')\big]}{c^2}E'=0
\label{eq:helmholtz_general},
\end{equation}
where $\omega_0' = c k_0'$ and $\omega_p^2(x') = 4 \pi e^2 n_e(x') / m_e \gamma$ is the local plasma frequency. $k_0'$ is the source pulse wavenumber in the vacuum. For a counter-propagating source pulse that is propagating in a background plasma, before the collision with the mirror (cf. figure~\ref{setup}) $\omega_0' = \gamma(\omega_0+\beta c k_b)=\gamma(1+\beta\eta)\omega_0$, and the source pulse electric field amplitude transforms as $E'=\gamma(1+\beta\eta)E$, where $k_b=\sqrt{\omega_0^2-\omega_{p0}^2}/c$ is the source pulse wavenumber in the background plasma and $\eta=\sqrt{1-\omega_{p0}^2/\omega_0^2}$ is the corresponding refractive index of the incoming source laser pulse in the background plasma.
For a flat-top density profile, and according to figure~\ref{setup},

\begin{equation}
\omega_p^2(x')=
\begin{cases}
\omega_{p0}^2 & x'<0\ \text{or}\ x'>l',\\
\omega_{\mathrm{p\max}}^2 & 0<x'<l',
\end{cases}
\label{eq:flat_plasma_frequencies}
\end{equation}
where $\omega_{\mathrm{p\max}} = \sqrt{4\pi e^2 n_{\mathrm{e\max}}/m_e \gamma}$ is the maximum plasma frequency, constant inside the flat-top mirror, and a Lorentz invariant~\cite{LAMPE.1978} (as is $\omega_{p0}$). For normal incidence,

\begin{equation}
E'(x')=
\begin{cases}
T'e^{-ik_b'x'} & x'<0,\\
A'e^{ik_{m}'x'}+B'e^{-ik_{m}'x'} & 0<x'<l',\\
e^{-ik_b'x'}+R'e^{ik_b'x'} & x'>l',
\end{cases}
\label{eq:piecewise_field}
\end{equation}
where $k_{m}'=\sqrt{\omega_0'^2-\omega_{\mathrm{p\max}}^{2}}/c$ is the source pulse wavenumber inside the mirror; $R' = E'_r / E'$ and $T' = E'_t/ E'$ are the amplitude reflection and transmission coefficients, and $k_b'=\sqrt{\omega_0'^2-\omega_{p0}^{2}}/c$. We observe that generalization to oblique incidence can be easily achieved by a Lorentz transformation in the transverse direction~\cite{gibbon2005short}. Using the state vector $\psi(x')$~\cite{born2013principles},

\begin{equation}
 \psi(x')=\begin{pmatrix}E'(x')\\[2pt]\dfrac{c}{\omega_0'}\,\partial_{x'}E'(x')\end{pmatrix},
\end{equation}
and evaluating $E'$ and $\partial_{x'}E'$ outside the mirror and dropping the propagation phase $e^{-2 i k_b' l'}$, which only shifts the phase of $R'$ and does not affect its magnitude,

\begin{equation}
 \psi(0^-)
=\begin{pmatrix}T'\\[4pt] -i\,\eta'\,T'\end{pmatrix},
\qquad
 \psi(l'^+)
=\begin{pmatrix}{1+R'}\\[4pt]
i\,\eta'\big(R'-1\big)
\end{pmatrix},
\end{equation}

while inside the mirror,
\begin{equation}
 \psi(0^+)
=\begin{pmatrix}
A'+B'\\[4pt]
i\,\chi'\eta'(A'-B')
\end{pmatrix},
\qquad
 \psi(l'^-)
=\begin{pmatrix}
A'e^{ik_{m}'l'}+B'e^{-ik_{m}'l'}\\[4pt]
i\,\chi'\eta'
\big(A'e^{ik_{m}'l'}-B'e^{-ik_{m}'l'}\big)
\end{pmatrix},
\end{equation}

where $\chi'=k_{m}'/k_b'$ and $\eta'=\sqrt{1-\omega_{p0}^2/\omega_0'^2}$. Continuity conditions $\psi(0^+)=\psi(0^-)$ and $\psi(l'^-)=\psi(l'^+)$ give

\begin{equation}
R'_{\mathrm{flat}}=\frac{(1-e^{-2i\theta})}{\big(\frac{1-\chi'}{1+\chi'}\big)-\big(\frac{1+\chi'}{1-\chi'}\big)e^{-2i\theta}},
\label{eq:Rprime_fraction_xi}
\end{equation}
with $\theta=k_{m}'l'$. Using $1-e^{-2i\theta}=2i\sin\theta\,e^{-i\theta}$, we readily obtain 

\begin{equation}
R'_{\mathrm{flat}}=\frac{2i\sin \theta(1-\chi'^2)}{(1-\chi')^2e^{i\theta}-(1+\chi')^2e^{-i\theta}},
\label{eq:Rprime_phaseful}
\end{equation}
substituting $e^{\pm i\theta}=\cos\theta\pm i\sin\theta$, and dividing both numerator and denominator by $\sin\theta$, yields
\begin{equation}
R'_{\mathrm{flat}}=\frac{i(1-\chi'^2)}{-2\chi'\cot\theta+i(1+\chi'^2)},
\label{eq:Rprime_phasefree}
\end{equation}
and using the plasma mirror dispersion relation to write $\chi' = \sqrt{(\omega_0'^2-\omega_{\mathrm{p\max}}^2)/(\omega_0'^2-\omega_{p0}^2)}$ (so that $\chi'\eta'=\sqrt{1-\omega_{\mathrm{p\max}}^2/\omega_0'^2}$, the corresponding refractive index of the incoming source laser pulse in the flat-top plasma mirror) and $\theta=l'\sqrt{\omega_0'^2-\omega_{\mathrm{p\max}}^2}/c$, we obtain

\begin{equation}
R'_{\mathrm{flat}}=\frac{\omega_{\mathrm{p\max}}^2-\omega_{p0}^2}{(2\omega_0'^2-\omega_{\mathrm{p\max}}^2-\omega_{p0}^2)
+2i\,\chi'\eta'^2\omega_0'^2\cot(\chi'\eta'k_0'l'_{\mathrm{mirror}})}.
\label{eq:Rprime_final_proper_flat}
\end{equation}

Equation~\eqref{eq:Rprime_final_proper_flat} gives the reflection coefficient of a counter-propagating source pulse interacting with a flat-top density profile mirror, both embedded in a background plasma, evaluated in the proper frame. The laboratory frame reflectivity is $|R| = |E_r/E| = \Gamma^2 |R'|$, where $E_r'= E_r / \gamma(1+\beta\eta)$ and  $\Gamma = \gamma(1+\eta\beta)$. With all quantities expressed in lab frame parameters, the reflection coefficient in the lab frame is 
\begin{equation}
\Bigl| R_{\mathrm{flat}} \Bigr|= \Gamma^2
\Biggl| \frac{\omega_{\mathrm{cut\text{-}off}}^{2}- \frac {\omega_{p0}^{2}}{\Gamma^2}}{
\!(2\omega_0^{2}-\omega_{\mathrm{cut\text{-}off}}^{2}-\frac {\omega_{p0}^{2}}{\Gamma^{2}})
+2i\,\chi \eta \sqrt{1-\frac{\omega_{p0}^2}{\Gamma^2\omega_0^2}}\,\omega_0^2
\cot\!\big(\chi \eta \gamma^2(1+\eta \beta)k_0l_{\mathrm{mirror}}\big)}\Biggr|,
\label{eq:R_final_lab_flat}
\end{equation}
where $\omega_{\mathrm{cut\text{-}off}}=\omega_{\mathrm{p\max}}/\Gamma$, $\chi=k_{m}/k_b$, $k_{m}=\sqrt{\omega_0^2-\omega_{\mathrm{cut\text{-}off}}^{2}}/c$, $\chi \eta=\sqrt{1-\omega_{\mathrm{cut\text{-}off}}^2/\omega_0^2}$, $l'_{\mathrm{mirror}}=\gamma l_{\text{mirror}}$, $k_0' = \Gamma k_0$ and $k_0=\omega_0/c$. In equation~\eqref{eq:R_final_lab_flat}, the periodic dependence associated with the $\cot\theta$ dependence accounts for the constructive/destructive interference due to the reflection on the front and on the back surface of the flat-top mirror.

We now consider an infinitely thin relativistic mirror, evaluating equation~\eqref{eq:R_final_lab_flat} in the $\delta$-Dirac limit. In this limit, $n'_e(x') = n_{\mathrm{mx}}\,\delta(x'-x_0')$, where $x_0'$ is the position of the density spike and the surface density, $\int n'_e(x')\, dx' = n'_{\mathrm{e\max}} l'_{\mathrm{mirror}} = n_{\mathrm{mx}}$, is constant. Starting from equation~\eqref{eq:Rprime_final_proper_flat} and taking $l'_{\mathrm{mirror}} \to 0$ at constant surface density, and with $n'_{\mathrm{e\max}}= n_{\mathrm{mx}}/l'_{\mathrm{mirror}}$, such that $n'_{\mathrm{e\max}} l'_{\mathrm{mirror}}=\mathrm{const}$, implies $\omega_0' \ll \omega_{\mathrm{p\max}}$, and one obtains $\chi'\eta' \simeq  i\kappa'$, 
$\kappa' \simeq {\omega_{\mathrm{p\max}}} / {\omega_0'}$, and the cotangent argument $\to i\,\omega_{\mathrm{p\max}} l'_{\mathrm{mirror}}/c$. Using $\omega_{\mathrm{p\max}}^2\propto n'_{\mathrm{e\max}}\propto 1/l'_{\mathrm{mirror}}$, it follows that $\omega_{\mathrm{p\max}} l'_{\mathrm{mirror}}/c \propto \sqrt{l'_{\mathrm{mirror}}} \to 0$. Thus equation~\eqref{eq:Rprime_final_proper_flat} reduces to

\begin{equation}
R'_{\mathrm{Dirac}}=-\frac{1-\frac{\omega_{p0}^2}{\omega_{\mathrm{p\max}}^2}}
{1+\frac{\omega_{p0}^2}{\omega_{\mathrm{p\max}}^2}-\frac{2i\,\eta'\omega_0'^2}{k_0'l'\omega_{\mathrm{p\max}}^2}}.
\label{eq:Rprime_final_proper_Dirac}
\end{equation}

Equation~\eqref{eq:Rprime_final_proper_Dirac} gives the mirror reflection coefficient for a $\delta$-Dirac limit in the proper frame. Using the same Lorentz transformation, $|R| = \Gamma^2 |R'|$, and rewriting all parameters in laboratory frame, we obtain
\begin{equation}
\Bigl| R_{\mathrm{Dirac}} \Bigr|=\Gamma^2 \Biggl|- \frac{1-\frac{\omega_{p0}^2}{\omega_{\mathrm{p\max}}^2}}
{1+\frac{\omega_{p0}^2}{\omega_{\mathrm{p\max}}^2}
-\frac{2i\sqrt{1-\frac{\omega_{p0}^2}{\Gamma^2\omega_0^2}}(1+\eta \beta)\omega_0^2}{k_0l\omega_{\mathrm{p\max}}^2}} \Biggr|.
\label{eq:R_final_lab_Dirac}
\end{equation}

\vspace{0.5\baselineskip}

In the present work, the Dirac-limit is modeled as an infinitely thin electron layer with finite areal density \(n_{\mathrm{mx}}=\int n'_e(x')\,dx'\), so that the proper-frame reflection coefficient is governed directly by this surface quantity. As \(n_{\mathrm{mx}}\) increases, the proper-frame field-amplitude reflection coefficient grows and approaches \(|R'|\to 1\), corresponding to a perfectly reflecting sheet in the proper frame. In contrast, Bulanov \textit{et al.}~\cite{bulanov2003light,kato2026high} describe the density spike near wave breaking as a \(\delta\)-like structure in a different physical regime, with reflectivity \(R \simeq 2\gamma\,(\omega_d/\omega_s)^2\), which remains partial. In that case, the intensity enhancement is dominated by Doppler upshift, pulse compression, and focusing. The present formulation therefore provides a complementary description in which the role of the areal density is explicit, and the transition from weak scattering to near-perfect reflection in the proper frame can be analyzed within a unified framework.

A similar technique as for the flat-top profile can be used for a parabolic density profile. According to figure~\ref{setup}, the plasma mirror frequency can be defined as

\begin{equation}
\omega_p^2(x')=
\begin{cases}
\omega_{p0}^2 & x'<0\ \text{or}\ x'>l',\\
\omega_{p0}^2+\frac{4\Delta^2}{l'^2}(x')^2 & 0<x'<\tfrac{l'}{2}, \\
\omega_{p0}^2+\frac{4\Delta^2}{l'^2}(x'-l')^2 & \tfrac{l'}{2}<x'<l',
\end{cases}
\label{eq:parabolic_plasma_frequencies}
\end{equation}
where $\Delta^2=\omega_{\mathrm{p\max}}^2-\omega_{p0}^2\ge0$. Equation~\eqref{eq:parabolic_plasma_frequencies} guarantees $\omega_p(0)=\omega_p(l')=\omega_{p0}$,  and reduces to the uniform background plasma frequency in the limit $\Delta\!\to\!0$. With the generalization $\omega_p^2(x')=\omega_{p0}^2+({4\Delta^2} / {l'^2})(x'-\epsilon l')^2$, where $\epsilon = 0, 1$ for left half and the right half of the mirror, respectively, and substituting into equation~\eqref{eq:helmholtz_general},
\begin{equation}
\frac{d^2E'}{dx'^2}
+\left[\frac{\omega_0'^2-\omega_{p0}^2}{c^2}
-\frac{4\Delta^2}{l'^2c^2}(x'-\epsilon l')^2\right]E'=0,
\label{eq:ode_plasma_x}
\end{equation}
can be reduced to the parabolic-cylinder form
\begin{equation}
\frac{d^2E'}{d\zeta'^2}+(\Lambda'-\zeta'^2)E'=0,
\label{eq:pcf}
\end{equation}
where $\zeta'^2 = ({2\Delta} / {l'c})\,(x'-\epsilon l')^2 $ and $\Lambda'=(\omega_0'^2-\omega_{p0}^2)l'/2c\Delta$. Assuming $E'(\zeta')=\sum_{n=0}^{\infty}a_n\zeta'^n$ yields $a_{m+2}=(a_{m-2}-\Lambda' a_m)/(m+2)(m+1)$, and separating for the even and odd branches~\cite{abramowitz1965handbook},

\begin{subequations}
\begin{equation}
E'(\zeta')=
e^{-\zeta'^{2}/2}\;{}_1F_1\!\Big(\tfrac{1-\Lambda'}{4},\tfrac12,\zeta'^{2}\Big)
\qquad \text{for the even branch}
\label{eq:even_field}
\end{equation}

\begin{equation}
E'(\zeta')=
\zeta'\, e^{-\zeta'^{2}/2}\;{}_1F_1\!\Big(\tfrac{3-\Lambda'}{4},\tfrac32,\zeta'^{2}\Big)
\qquad \text{for the odd branch}
\label{eq:odd_field}
\end{equation}
\end{subequations}
where ${}_1F_1$ is the Kummer confluent hypergeometric function~\cite{abramowitz1965handbook}. Applying the Kummer transformation ${}_1F_1(a,b,z)=e^{z}\,{}_1F_1(b-a,b,-z)$, the equations~\eqref{eq:even_field} and~\eqref{eq:odd_field} can be written as

\begin{subequations}
\begin{equation}
E'(\zeta')=
e^{-\nu'_{\epsilon}/2}\,{}_1F_1\!\Big(\tfrac{\mu'_+}{4},\tfrac12,\nu'_{\epsilon}\Big)
\qquad \text{for the even branch}
\label{eq:even__field}
\end{equation}

\begin{equation}
E'(\zeta')=
(\pm i\sqrt{\nu'_{\epsilon}})\,e^{-\nu'_{\epsilon}/2}\,
{}_1F_1\!\Big(\tfrac{\mu'_++2}{4},\tfrac32,\nu'_{\epsilon}\Big)
\qquad \text{for the odd branch}
\label{eq:odd__field}
\end{equation}
\end{subequations}
where $\nu'_{\epsilon} = - \zeta'^2$ and $\mu'_\pm = 1\pm\Lambda' = 1\pm {(\omega_0'^2-\omega_{p0}^2)\,l'} / {2c\,\Delta}$. The electric field inside the mirror for left half and right half can be written as the summation of the even and odd results as equations~\eqref{eq:even__field} and~\eqref{eq:odd__field}. For normal incidence, 

\begin{equation}
E'(x')=
\begin{cases}
T'\,e^{-ik_b' x'} & x'<0,\\[6pt]

e^{-\nu'_0/2}\left[
A'\,{}_1F_1\!\Big(\tfrac{\mu'_+}{4},\tfrac12,\nu'_0\Big)
+
B'\,\sqrt{\nu'_0}\,{}_1F_1\!\Big(\tfrac{\mu'_++2}{4},\tfrac32,\nu'_0\Big)
\right]
& 0<x'<\tfrac{l'}{2},\\[8pt]

e^{-\nu'_1/2}\left[
C'\,{}_1F_1\!\Big(\tfrac{\mu'_+}{4},\tfrac12,\nu'_1\Big)
+
D'\,\sqrt{\nu'_1}\,{}_1F_1\!\Big(\tfrac{\mu'_++2}{4},\tfrac32,\nu'_1\Big)
\right]
& \tfrac{l'}{2}<x'<l',\\[8pt]

e^{-ik_b' x'}+R'\,e^{ik_b' x'} & x'>l', 
\label{eq:electric__field}
\end{cases}
\end{equation}
 where $\pm i$ are absorbed into constants. Using the state vector~\cite{born2013principles}, and redefining $\psi$ with the normalization $c/\Delta$ for the parabolic case,

\begin{equation}
 \psi(x')=\begin{pmatrix}E'(x')\\[2pt]\dfrac{c}{\Delta}\,\partial_{x'}E'(x')\end{pmatrix},
\end{equation}
evaluating $E'$ and $\partial_{x'}E'$ in the background plasma on either side of the mirror and ignoring propagation phase $e^{-2 i k_b' l'}$, 
\begin{equation}
 \psi(0^-)
=\begin{pmatrix}T'\\ -i\,\dfrac{\omega_0' \eta'}{\Delta}\,T'\end{pmatrix},
\qquad
 \psi(l'^+)
=\begin{pmatrix}1+R'\\ i\,\dfrac{\omega_0' \eta'}{\Delta}\,(R'-1)\end{pmatrix}, 
\end{equation}

while inside the mirror, using the identity $\partial_{z}\,{}_1F_1(a,b,z)= ({a} / {b})\,{}_1F_1(a+1,b+1,z)$, where $b\neq 0,-1,-2,\dots$, 

\begin{equation}
\psi(0^+)=
\begin{pmatrix}
A' \\
-\sqrt{\dfrac{-2c}{\Delta l'}}\,B'
\end{pmatrix},
\qquad
\psi(l'^-)=
\begin{pmatrix}
C' \\
\sqrt{\dfrac{-2c}{\Delta l'}}\,D'
\end{pmatrix},
\end{equation}
and at the midpoint of the parabolic mirror ($x'=  {l'} / {2}$), one has $\nu'_0=\nu'_1 = \nu'_m$, the states are 

\begin{subequations}
\begin{equation}
\psi\!\left(\tfrac{l'}{2}^-\right)=
\begin{pmatrix}
e^{-\nu'_m/2}\left[
A'\,{}_1F_1\!\Big(\tfrac{\mu'_+}{4},\tfrac12,\nu'_m\Big)
+
B'\,\sqrt{\nu'_m}\,{}_1F_1\!\Big(\tfrac{\mu'_++2}{4},\tfrac32,\nu'_m\Big)
\right]\\[8pt]
e^{-\nu'_m/2}
\Bigg[
A'\,{}_1F_1\!\Big(\tfrac{\mu'_+}{4},\tfrac12,\nu'_m\Big)
+
B'\sqrt{\nu'_m}\,{}_1F_1\!\Big(\tfrac{\mu'_++2}{4},\tfrac32,\nu'_m\Big)
\\[8pt] - A'\mu'_+\,
{}_1F_1\!\Big(\tfrac{\mu'_++4}{4},\tfrac32,\nu'_m\Big)
-
\frac{B'}{\sqrt{\nu'_m}}\,
{}_1F_1\!\Big(\tfrac{\mu'_++2}{4},\tfrac32,\nu'_m\Big)
-
\frac{B'(\mu'_++2)}{3}\sqrt{\nu'_m}\,
{}_1F_1\!\Big(\tfrac{\mu'_++6}{4},\tfrac52,\nu'_m\Big)
\Bigg]
\end{pmatrix},
\end{equation}

\begin{equation}
\psi\!\left(\tfrac{l'}{2}^+\right)=
\begin{pmatrix}
e^{-\nu'_m/2}\left[
C'\,{}_1F_1\!\Big(\tfrac{\mu'_+}{4},\tfrac12,\nu'_m\Big)
+
D'\,\sqrt{\nu'_m}\,{}_1F_1\!\Big(\tfrac{\mu'_++2}{4},\tfrac32,\nu'_m\Big)
\right]\\[8pt]
-e^{-\nu'_{m}/2}
\Bigg[
C'\,{}_1F_1\!\left(\frac{\mu'_+}{4},\frac{1}{2},\nu'_{m}\right)
+
D'\sqrt{\nu'_{m}}\,{}_1F_1\!\left(\frac{\mu'_++2}{4},\frac{3}{2},\nu'_{m}\right)
\\[8pt]-
C'\mu'_+\,{}_1F_1\!\left(\frac{\mu'_++4}{4},\frac{3}{2},\nu'_{m}\right)
-
\frac{D'}{\sqrt{\nu'_{m}}}\,{}_1F_1\!\left(\frac{\mu'_++2}{4},\frac{3}{2},\nu'_{m}\right)
-
\frac{D'(\mu'_++2)}{3}\sqrt{\nu'_{m}}\,
{}_1F_1\!\left(\frac{\mu'_++6}{4},\frac{5}{2},\nu'_{m}\right)
\Bigg]
\end{pmatrix}.
\end{equation}
\end{subequations} 

With the continuity conditions $\psi(0^-)= \psi(0^+)$, $\psi\!\left(\frac{l'}{2}^{-}\right)= \psi\!\left(\frac{l'}{2}^{+}\right)$, and $\psi(l'^-)= \psi(l'^+)$, and using the identity ${}_1F_1(a+1,b+1,z)= ({b} / {a z}) \left[{}_1F_1(a,b+1,z)-{}_1F_1(a,b,z)\right]$, the reflection coefficient for a parabolic mirror in the proper frame is

\begin{equation}
\begin{aligned}
R'_{\mathrm{parabolic}} \;=\; 1
&-\frac{{}_1F_1\!\Big(\frac{\mu'_+}{4}, \frac{1}{2}, \nu'_m\Big)}
{{}_1F_1\!\Big(\frac{\mu'_+}{4}, \frac{1}{2}, \nu'_m\Big)-i\,\frac{l'k_b'}{2}\;
{}_1F_1\!\Big(\frac{\mu'_+ + 2}{4}, \frac{3}{2}, \nu'_m\Big)}
\\[8pt]
&\hspace{-8.0em}+ \frac{{}_1F_1\!\Big(\frac{\mu'_+}{4}, \frac{1}{2}, \nu'_m\Big)-\mu'_+\,{}_1F_1\!\Big(\frac{\mu'_+}{4}+1, \frac{3}{2}, \nu'_m\Big)}
{\mu'_+\,{}_1F_1\!\Big(\frac{\mu'_+}{4}+1, \frac{3}{2}, \nu'_m\Big)-{}_1F_1\!\Big(\frac{\mu'_+}{4}, \frac{1}{2}, \nu'_m\Big)
+i\!\left[\frac{l'k_b'}{2}\;{}_1F_1\!\Big(\frac{\mu'_+ + 2}{4}, \frac{3}{2}, \nu'_m\Big)
+\frac{\omega_0' \eta'}{\Delta}\;{}_1F_1\!\Big(\frac{\mu'_+ + 2}{4}, \frac{1}{2}, \nu'_m\Big)\right]} \,.
\end{aligned}
\label{eq:R_final_proper_parabolic}
\end{equation} 

Equation~\eqref{eq:R_final_proper_parabolic} gives the reflection coefficient for a parabolic density profile mirror in the proper frame. After multiplying $R'_{\mathrm{parabolic}}$ by $\Gamma^2 $ and converting all quantities to the laboratory frame, as for the flat-top and Dirac-delta limit, the parabolic reflection coefficient is given by

\begin{equation}
\begin{aligned}
\Bigl |R_{\mathrm{parabolic}} \Bigr|= \Gamma^2
\Biggl|
1
&-
\frac{
{}_1F_1\!\Big(\frac{\mu_+}{4}, \frac{1}{2}, \nu_m\Big)}{
{}_1F_1\!\Big(\frac{\mu_+}{4}, \frac{1}{2}, \nu_m\Big)
-
i\,\frac{\gamma l}{2}
\frac{\sqrt{(\Gamma\omega_0)^2-\omega_{p0}^2}}{c}\,
{}_1F_1\!\left(\frac{\mu_+ + 2}{4},\frac32, \nu_m\right)}
\\[8pt]
&\hspace{-8.0em}+ \frac{
{}_1F_1\!\Big(\frac{\mu_+}{4}, \frac{1}{2}, \nu_m\Big)
-
\mu_+\,
{}_1F_1\!\left(\frac{\mu_+}{4}+1,\frac32, \nu_m\right)
}{
\begin{aligned}
&\mu_+\,
{}_1F_1\!\Big(\frac{\mu_+}{4}+1,\frac32, \nu_m\Big)
-
{}_1F_1\!\Big(\frac{\mu_+}{4}, \frac{1}{2}, \nu_m\Big)
\\
&\quad +
i\left[
\frac{\gamma l}{2}
\frac{\sqrt{(\Gamma\omega_0)^2-\omega_{p0}^2}}{c}\,
{}_1F_1\!\Big(\frac{\mu_+ + 2}{4},\frac32, \nu_m\Big)
+
\frac{\Gamma\omega_0}{\Delta}
\sqrt{1-\left(\frac{\omega_{p0}}{\Gamma\omega_0}\right)^{2}}\,
{}_1F_1\!\Big(\frac{\mu_+ + 2}{4},\frac12, \nu_m\Big)
\right]
\end{aligned}
}
\Biggr| \,,
\end{aligned}
\label{eq:R_final_lab_parabolic}
\end{equation}
where $\mu_+ = \mu'_+$, $\Lambda = \Lambda' = {\left((\Gamma\omega_0)^2-\omega_{p0}^2 \right)\gamma l} / {2c\Delta}$, and 
$\nu_m = \nu'_m = - {\Delta\,\gamma l} / {2c}$. 
Finally, the transmission coefficient is obtained from energy conservation in the proper frame, $|T'|^{2}+|R'|^{2}=1$, and transforms as $|T| = |E_t/E| = |T'|$, since $E_t'=\gamma(1+\beta\eta)E_t$ for a counter-propagating source pulse in the background plasma. 

While equation~\eqref{eq:R_final_lab_parabolic} is a bit cumbersome to draw conclusions from its inspection, the numerical evaluation and comparison between equations~\eqref{eq:R_final_lab_flat}, \eqref{eq:R_final_lab_Dirac}, and~\eqref{eq:R_final_lab_parabolic} reveal the strong deviations associated with non-ideal relativistic mirrors.

\begin{figure}
\centering

\includegraphics[width=0.63\linewidth,height=5.73cm]{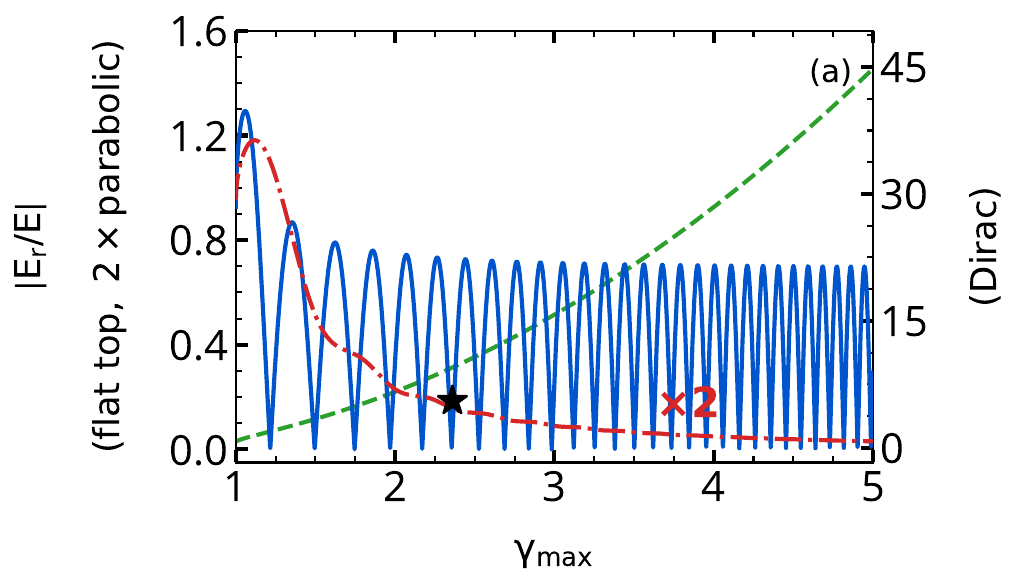}

\includegraphics[width=0.63\linewidth,height=5.73cm]{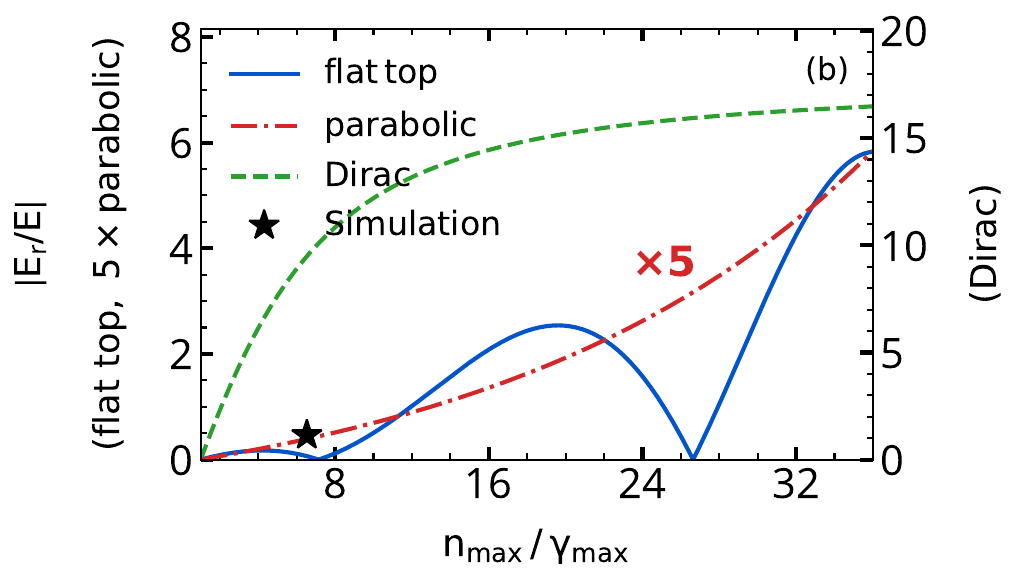}\\[6pt]

\includegraphics[width=0.63\linewidth,height=5.73cm]{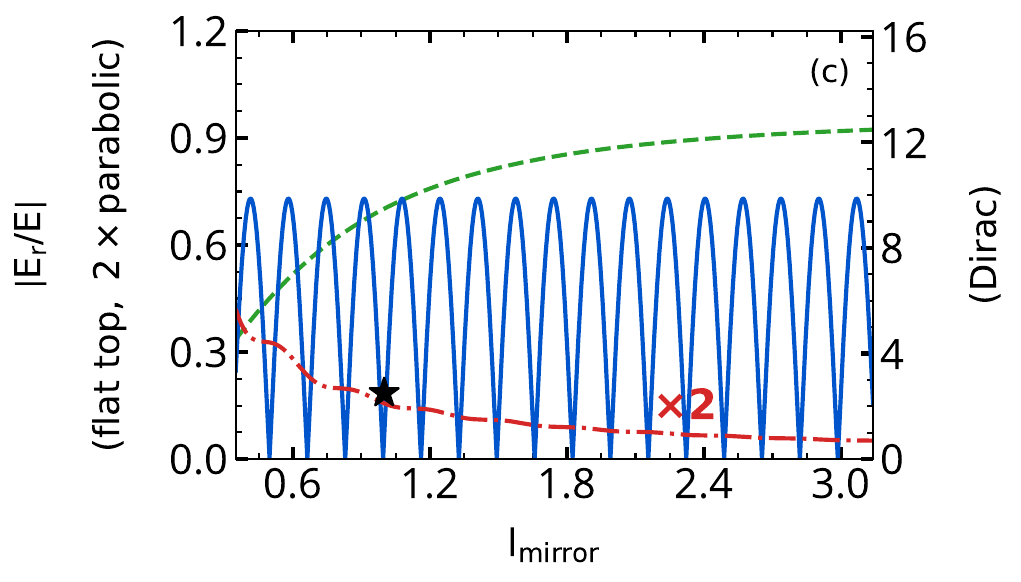}\\[6pt]
\caption{The magnitude of the reflection coefficient, $|R|=|E_r/E|$, for flat-top (solid blue), parabolic (dash-dotted red), and Dirac-limit (dashed green) density profiles as a function of (a) $\gamma_{\max}$, (b) $n_{\mathrm{e\max}}/\gamma_{\max}$, and (c) $l_{\mathrm{mirror}}$. In each panel, one parameter is varied while the others are fixed to $\gamma_{\max}=2.36$, $n_{\mathrm{e\max}}/\gamma_{\max}=6.53$, and $l_{\mathrm{mirror}}=1.00 \, c/\omega_{p0}$, corresponding to the values obtained from simulations at $a_0=3.0$ ($\omega_0=2\omega_{p0}$).
The flat-top profile exhibits strong oscillations due to interference between reflections at its two sharp boundaries, while these oscillations are suppressed for the parabolic profile due to its smooth density gradient. In contrast, in the Dirac limit, reflectivity is governed by the areal density, and interference effects are absent. In panel (a), both $|R_{\mathrm{parabolic}}|$ and the oscillation envelope of $|R_{\mathrm{flat}}|$ decrease with increasing $\gamma_{\max}$ as the mirror approaches relativistic transparency; in panel (b), they increase with $n_{\mathrm{e\max}}/\gamma_{\max}$ as the mirror becomes increasingly overdense. In panel (c), increasing $l_{\mathrm{mirror}}$ shifts the flat-top oscillation pattern without changing its peak reflectivity and reduces $|R_{\mathrm{parabolic}}|$ as the refractive-index variation becomes more adiabatic. Across all panels, however, $|R_{\mathrm{Dirac}}|$ increases monotonically because increasing each parameter raises the areal density, driving the mirror toward the perfectly reflecting-sheet limit.
Black stars denote the corresponding \textsc{Osiris} simulation results, which closely follow the parabolic model prediction and are plotted with the same scaling factors as the curves in each panel.}
\label{Rsngl}
\end{figure}

\section{Reflection coefficient of laser-driven plasma mirrors and PIC simulations}

To understand the impact of the deviations to non-ideal conditions of relativistic mirrors, we now explore the dependence of the reflection coefficient on physical parameters of the mirror, namely $n_{\mathrm{e\max}}$, $\gamma_{\max}$, and $l_{\mathrm{mirror}}$. For the baseline conditions, we employ the parameters from a 2D \textsc{Osiris} simulation (parameters described in the supplementary material) for $a_0=3.0$, and a pulse duration $15\,\mathrm{fs}$, which drives a relativistic mirror with $n_{\mathrm{e\max}}/\gamma_{\max}=6.53$, $\gamma_{\max}=2.36$, and $l_{\mathrm{mirror}}=1.00 \, c/\omega_{p0}$. The reflectivity measured directly in the simulations is also presented for comparison. It is evaluated from an on-axis lineout as the ratio of the peak envelopes of the reflected and incident source electric fields near the mirror, excluding diffraction and focusing effects.

We now analyze the reflection coefficients for the three density profiles discussed in this paper—flat-top, Dirac-$\delta$, and parabolic—given by equations~\eqref{eq:R_final_lab_flat}, \eqref{eq:R_final_lab_Dirac}, and \eqref{eq:R_final_lab_parabolic}. 

For the flat-top profile (blue curves) in figures~\ref{Rsngl}(a)--(c), $|R_{\mathrm{flat}}|$ exhibits oscillations due to the interference between reflections at the two sharp density discontinuities at the mirror boundaries. In figure~\ref{Rsngl}(a), the oscillation envelope decreases as the mirror approaches relativistic transparency, consistent with the progressive reduction in reflectivity as $\Gamma^2\omega_0^2$ approaches and exceeds $\omega_{\mathrm{p\max}}^2$. In figure~\ref{Rsngl}(b), the oscillation envelope of $|R_{\mathrm{flat}}|$ increases with $n_{\mathrm{e\max}}/\gamma_{\max}$ as the mirror becomes increasingly overdense in its proper frame. Figure~\ref{Rsngl}(c) shows that at fixed proper density and Lorentz factor, varying $l_{\text{mirror}}$ changes the interference phase, $\theta=k'_m l'$, and therefore shifts the oscillation pattern, while leaving the peak reflectivity unchanged.

For the parabolic density profile (red curves) in figures~\ref{Rsngl}(a)--(c), the oscillations are also strongly suppressed: the mirror density rises and falls smoothly, reflection occurs along the gradient rather than at two abrupt interfaces, thus suppressing the interference between reflections from the two surfaces of the flat-top mirror. In figures~\ref{Rsngl}(a) and~\ref{Rsngl}(b) the trends with $\gamma_{\max}$ and $n_{\mathrm{e\max}}/\gamma_{\max}$ follow the same physical interpretation as for the flat-top profile. In figure~\ref{Rsngl}(c), increasing $l_{\mathrm{mirror}}$ at fixed $n_{\mathrm{e\max}}/\gamma_{\max}$ and $\gamma_{\max}$ reduces the density gradient and makes the refractive-index variation more adiabatic, so reflected contributions along the gradient partially cancel, lowering $|R_{\mathrm{parabolic}}|$ over the range shown.

In the Dirac-$\delta$ limit (green curves) in figures~\ref{Rsngl}(a)--(c), the mirror is modeled as an infinitesimally thin sheet with finite surface density. The interaction is therefore controlled by the areal charge, and interference effects vanish. 
In this limit, $\omega_{\mathrm{p\max}}\to\infty$ and $\Gamma^2\omega_0^2\ll\omega_{\mathrm{p\max}}^2$, so the mirror remains strongly overdense and $|R_{\mathrm{Dirac}}|$ increases monotonically toward the perfectly reflecting-sheet limit, as shown in figures~\ref{Rsngl}(a)--(c). These trends follow from the corresponding increase in the areal density, $n_{\mathrm{mx}}=\int n'_e(x')\, dx' = n'_{\mathrm{e\max}}l'_{\text{mirror}}$, where $n'_{\mathrm{e\max}}=n_{\mathrm{e\max}}/\gamma_{\max}$ and $l'_{\mathrm{mirror}}=\gamma_{\max}l_{\mathrm{mirror}}$.

The \textsc{Osiris} results (black stars) in figures~\ref{Rsngl}(a)--(c) at $a_0=3.0$ are very close to the reflectivity predicted by the parabolic model, consistent with the driven density spike being well described by a parabolic profile.

\begin{figure}[H]
 \centering
\includegraphics[width=0.59\linewidth,height=6.8cm]{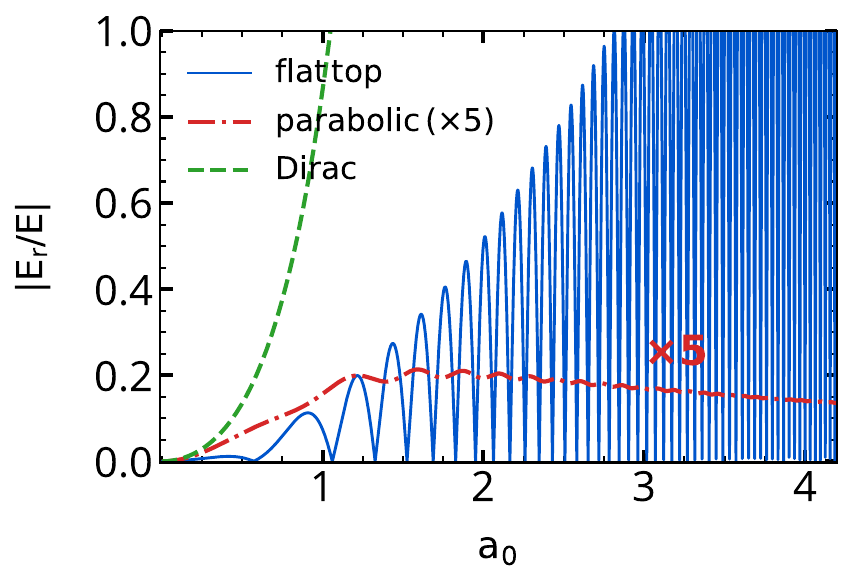}
\caption{The magnitude of the reflection coefficient, $|R|=|E_r/E|$, versus $a_0$ for flat-top (solid blue), parabolic (dash-dotted red), and Dirac (dashed green) density profiles, with $\omega_0 = 2\omega_{p0}$. The oscillations in $|R_{\mathrm{flat}}|$ arise from interference between reflections at two sharp interfaces, whereas the smooth dependence of $|R_{\mathrm{parabolic}}|$ results from reflection over a gradual density gradient. The Dirac profile shows no oscillations because reflection occurs at a single interface. For the flat-top profile, the envelope grows as $(\omega_{\mathrm{p\max}}^2-\omega_{p0}^2)/\omega_0^2\propto a_0^2$, while $|R_{\mathrm{Dirac}}|$ increases with the areal density, $n_{\mathrm{mx}}\propto a_0^3$ (for $a_0>1$). For the parabolic profile, the reflectivity reaches a maximum at $a_0\simeq 1.6$, set by the balance between electron compression, finite mirror thickness, and the onset of relativistic transparency. 
}
\label{RsTheory}
\end{figure}

In figure~\ref{RsTheory} we explore $|R|=|E_r/E|$ assuming the mirror parameters are determined by the quasi-static model, equations~\eqref{nemax} to~\eqref{lmirror}. This allows us to write the reflection coefficients, equations~\eqref{eq:R_final_lab_flat}, \eqref{eq:R_final_lab_Dirac}, and \eqref{eq:R_final_lab_parabolic} as a function of $a_0$ only and determine the optimal $a_0$ to drive a relativistic mirror.  We explore the three density models. 
As before, the flat-top result displays strong oscillations due to the two sharp discontinuities of the density profile, whereas the parabolic profile yields a smooth dependence due to the density gradient. The Dirac-$\delta$ limit shows no oscillations because reflection occurs at a single effective interface. For the flat-top profile, the combined effects of $n_{\mathrm{e\max}}/\gamma_{\max}$ and the laboratory-frame transformation, $|R|=\Gamma^2|R'|$, yield the envelope scaling $(\omega_{\mathrm{p\max}}^2-\omega_{p0}^2)/\omega_0^2\propto a_0^2$, whereas $|R_{\mathrm{Dirac}}|$ increases with the areal density, $n_{\mathrm{mx}}\propto a_0^3$, in our working regime ($a_0>1$).

For the parabolic model, $|R_{\mathrm{parabolic}}|$ reaches a maximum at $a_0\simeq 1.6$, set by the competition between increasing electron compression, its thickness, and the approach to relativistic transparency ($\omega_0 = 2\omega_{p0}$). The reflection coefficient is higher at lower frequencies. As follows from~\eqref{eq:R_final_lab_flat}, \eqref{eq:R_final_lab_Dirac}, and~\eqref{eq:R_final_lab_parabolic}, the reflectivity exhibits a strong dependence on the source pulse frequency, in addition to $n_{\mathrm{e\max}}$, $\gamma_{\max}$, and $l_{\mathrm{mirror}}$.

\section{Conclusions}

Analytical expressions for the reflection coefficient of relativistic plasma mirrors driven by intense laser pulses in underdense plasmas were derived for three representative density profiles—flat-top, parabolic, and Dirac-limit. This allowed us to isolate the role of the mirror density shape on reflectivity. 2D \textsc{Osiris} simulations and our analytical model confirm that the parabolic density profile provides a good approximation to the density spike formed at the rear of the plasma bubble. 

For the parabolic density profile of the mirror and under the quasi-static approximation and assuming a step-like laser driver, the reflectivity attains an optimum at a specific laser amplitude, set by the balance between the relativistic plasma mirror proper density, thickness, and its relativistic factor, whereas the flat-top and Dirac-limit density profiles of the mirror remain monotonically increasing in the explored regime. These results clarify how the reflection efficiency depends on the plasma mirror parameters and establish a consistent theoretical framework for interpreting and optimizing relativistic plasma-mirror schemes for efficient ultrashort pulse generation. Future work will examine the impact of polarization of the driver on mirror formation and reflectivity, and the role of curved relativistic mirrors in producing a flying focus.

%
% Each of the commands below will create an unnumbered section with the appropriate heading.
% Remove any sections that are not relevant for your article.
% All sections except suppdata will be removed if the [anonymous] option is used.
% See iopjournal-guidelines.pdf for more information.
%

\ack{We acknowledge useful discussions with Dr.\ T.~Silva and the initial calculations on the reflectivity by Dr.\ J.~Santos and Dr.\ S.~Martins that have inspired our work.

The authors declare that generative artificial intelligence tools were used exclusively to support the linguistic and stylistic revision of the final version of the text, and that the scientific content and derivations are entirely their own responsibility.}

\funding{This work was supported by FCT (Portugal) through the PhD grant 2025.02146.BD and the project X-MASER (2022.02230.PTDC).
Simulations were performed at Deucalion (Portugal) funded by FCT Masers in Astrophysical Plasmas 2 (MAPs2) I.P project 2024.11062.CPCA.A3 (doi:10.54499/2023.11062.CPCA.A3); and at MareNostrum 5 (Spain) through the Project 2024.07895.CPCA.A3 (doi:10.54499/2024.07895.CPCA.A3).}
% This section is a list of funder names and grant numbers

\roles{Z.M.C. performed the simulations and their analysis, and developed the analytical model. L.O.S. provided guidance and supervision. Both authors contributed to the writing of the manuscript.}
% List author names and the contributions made to the article, using terms from the NISO Contributor Roles Taxonomy (CRediT) https://credit.niso.org

\data{The data that support the findings of this study are openly available at the following DOI: \url{https://doi.org/10.5281/zenodo.22845723}.}
% For more information on IOP Publishing's research data policy see: https://publishingsupport.iopscience.iop.org/questions/research-data/

\suppdata{Several 2D particle-in-cell \textsc{Osiris} simulations were performed for this work. The driver wavelength was $\lambda_{d}=0.8\,\mu\mathrm{m}$, with dimensionless amplitude $a_0=[1.0,\,1.5,\,2.0,\,2.5,\,3.0]$. The source pulse had $a_{s}=0.03$ and $\lambda_s = 2.86\lambda_d$. 

The polynomial temporal envelope used was $a_d(\xi)=a_0[10(\xi/L_d)^3-15(\xi/L_d)^4+6(\xi/L_d)^5]$ for \(0<\xi<L_d\), symmetric about \(\xi=L_d\), with $\xi=k_{p0}(x-ct)$. In all simulations, the transverse profiles were Gaussian.
The pulse durations and spot sizes were \(\tau_s = 2\tau_d = 11.2T_d\) and \(W_d = W_s = 8.44\lambda_d\), respectively, where \(T_d\) is the driver period.
The plasma slab of size $L_x \times L_y = 63.7 \times 31.8\lambda_d^2$ was embedded in a simulation box of $108.76\times 31.8\lambda_d^2$. It consisted of electrons with linear density ramps at both boundaries. The entrance ramp $31.85\lambda_d$ enhanced mirror symmetry and lifetime, while the exit ramp $8.0\lambda_d$ enabled a clean separation of the reflected pulse from the mirror. The plateau density was $n_e = 0.03\,n_{cr}$, where $n_{cr} = m_e\omega_d^2/(4\pi e^2)$. The spatial resolution was \(\Delta x = \lambda_d/60\) and \(\Delta y = \lambda_d/30\), with \(8\times 8\) super-particles per cell. Open boundary conditions were used for fields and particles, and the time step was \(\Delta t \approx T_d/320\). A \(y\)-polarized driver entered from the left boundary, while a \(z\)-polarized source entered from the right, both propagating along \(x\).}

\section*{}
\bibliography{references}
\bibliographystyle{iopart-num}

\end{document}